\documentclass[a4paper]{JAC2003}
\usepackage{graphicx}
\usepackage{amsmath}
\usepackage{amssymb}
\usepackage{amsfonts}
\usepackage{cite}

\begin{document}
\title{STATUS OF HEAD-ON BEAM--BEAM COMPENSATION IN RHIC
\thanks{Work supported by Brookhaven Science Associates LLC under Contract
no.~DE-AC02-98CH10886 with the U.S.\ Department of Energy.}}
\author{W. Fischer\thanks{Wolfram.Fischer@bnl.gov},
Z. Altinbas, M. Anerella, M. Blaskiewicz, D. Bruno, M. Costanzo, W.C. Dawson, \\
D.M. Gassner, X. Gu, R.C. Gupta, K. Hamdi, J. Hock, L.T. Hoff, R. Hulsart,
A.K. Jain, \\ R. Lambiase, Y. Luo,
M. Mapes, A. Marone, R. Michnoff, T.A. Miller, M. Minty, C. Montag, \\
J. Muratore, S. Nemesure, D. Phillips, A.I. Pikin, S.R. Plate, P. Rosas,
L. Snydstrup, Y. Tan, \\
C. Theisen, P. Thieberger, J. Tuozzolo, P. Wanderer, S.M. White, W. Zhang \\
BNL, Upton, NY, USA}

\maketitle

\begin{abstract}
In polarized proton operation, the performance of the
Relativistic Heavy Ion Collider (RHIC)  is limited by the head-on
beam--beam effect. To overcome this limitation, two electron lenses are under
commissioning. We give an overview of head-on beam--beam compensation in
general and in the specific design for RHIC, which is based on electron lenses.
The status of installation and commissioning are presented along with plans for
the future.
\end{abstract}

\section{Introduction}
Head-on beam--beam compensation was first  proposed as a four-beam
e$^+$e$^-$e$^+$e$^-$ scheme for COPPELIA~\cite{DCI0} and implemented for
Dispositif de Collisions dans l'Igloo (DCI)~\cite{DCI1}. The DCI experience, however, fell short of expectations;
luminosities with two, three, or four beams were about the same. The shortfall is generally
attributed to coherent beam--beam instabilities~\cite{DCI2,DCI3,DCI4}, and
head-on beam--beam compensation has not been tested again since.

Nevertheless, various proposals have been made,  such as for the SSC~\cite{Tsyg0,Tsyg1},
Tevatron~\cite{TEL3}, LHC~\cite{Tsyg0,Tsyg1,Tsyg2,Scan,Dord}, and
B-factories~\cite{Ohni}. In hadron colliders, the compensation can be achieved by
colliding positively charged beams with a negatively charged low-energy
electron beam, in a device usually referred to as an electron lens. Doing so avoids
the coherent instabilities seen in DCI, as the electron beam will not couple
back to the hadron beam, except for single-pass effects; these can be
significant~\cite{TEL2,Whit1} and may require the addition of a transverse
damper in RHIC. Two electron lenses
were installed in the Tevatron~\cite{TEL1,TEL2,TEL3,TEL4,TEL5,TEL6,TEL7},
where they were routinely used as a gap cleaner, but not for head-on beam--beam
compensation. The Tevatron experience is
valuable for several  reasons: (i) the reliability of the technology was
demonstrated, as no store was ever lost because of the lenses~\cite{Shil}; (ii)
the tune shift of selected bunches due to PACMAN effects was corrected,
leading to lifetime improvements~\cite{TEL4}; (iii) the sensitivity to
positioning errors, transverse profile shape, and electron beam current
fluctuations was explored~\cite{Stan}; (iv) experiments with a Gaussian
 profile electron beam were performed; and (v) a hollow electron beam was tested in a collimation
scheme~\cite{TEL7}. For the design of the RHIC electron lenses we have
benefited greatly from the Tevatron experience. We have also drawn on the
expertise gained in the construction and operation of an Electron Beam Ion
Source (EBIS) at Brookhaven National Laboratory (BNL)~\cite{EBIS1,EBIS2}, which is a device similar to an electron lens
but with a different purpose.

\begin{figure}[h]
\centering
\includegraphics[width=70mm]{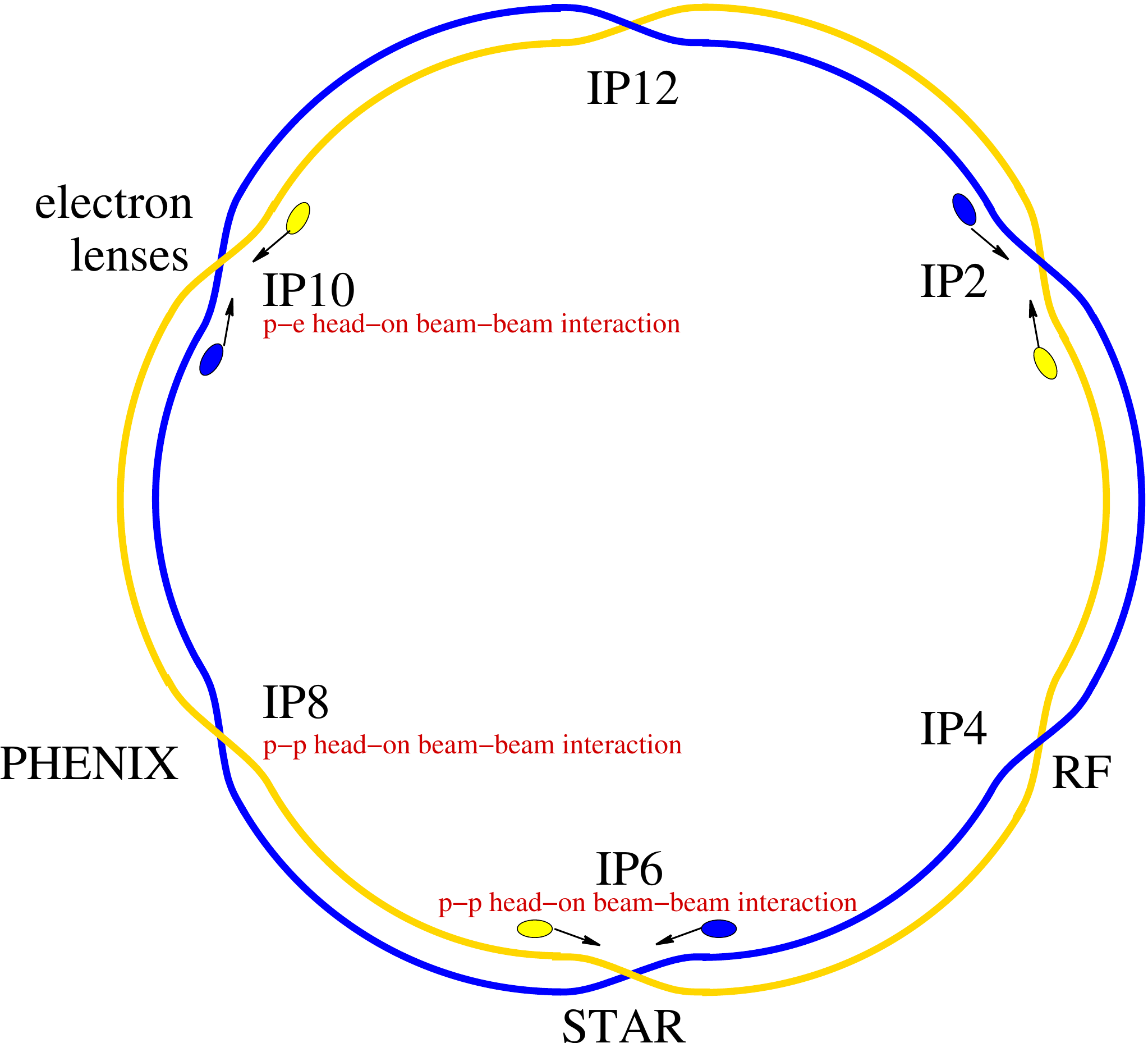}
\caption{General layout of RHIC with locations of the head-on beam--beam
interactions and electron lenses.}
\label{fig:ov}
\end{figure}

\begin{figure}[h]
\centering
\includegraphics[width=80mm]{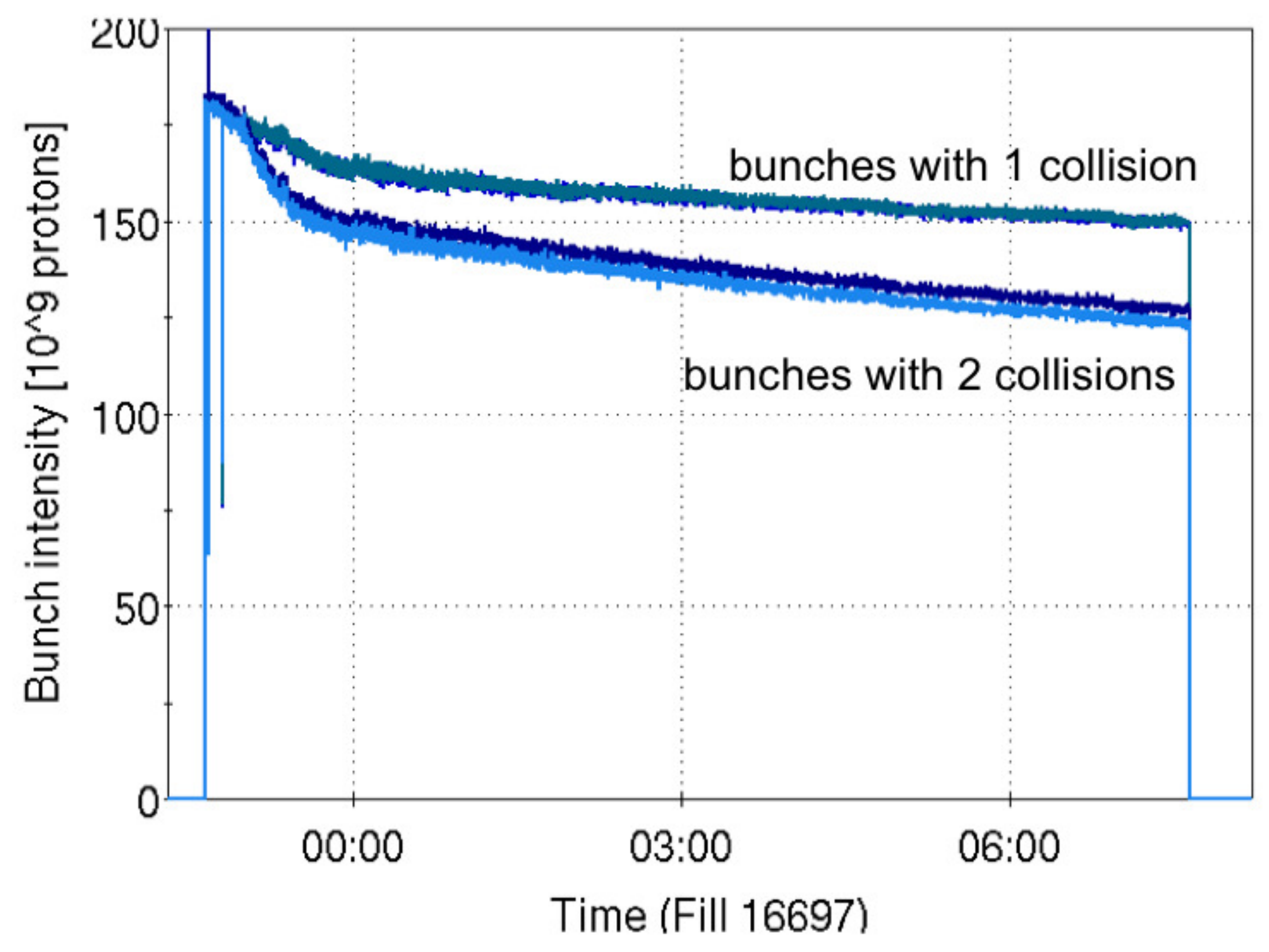}
\caption{Time-dependent intensity of polarized proton bunches with one or
two head-on collisions during the 2012 run.}
\label{fig:bunch12}
\end{figure}

In RHIC there are two head-on beam--beam interactions at interaction points IP6
and IP8 (Fig.~\ref{fig:ov}), as well as four long-range beam--beam interactions with large
separation (about 10~mm) between the beams at the other interaction points.
The luminosity
is limited by the head-on effect in polarized proton operation~\cite{RHICbb0,
RHICbb1,RHICbb2,RHICbb3,RHICbb4,RHICbb5,RHICbb6}, as can be seen in
Fig.~\ref{fig:bunch12}. Bunches with two collisions experience a larger proton
loss throughout the store than bunches with only one collision. The enhanced
loss is particularly strong at the beginning of a store. Beam--beam effects
in other hadron colliders are reported in Refs.~\cite{BB99,BB01,BB03,ICFA,
Shil3,BB13}.

We consider the partial indirect compensation of the head-on beam--beam effect
with one electron lens in each ring. Together with intensity and emittance
upgrades~\cite{Zele}, our goal is to approximately double the luminosity
over what can be achieved without these upgrades.

This article gives a summary of previous studies and progress reports on
 head-on beam--beam compensation in RHIC with electron lenses~\cite{Luo0,Luo1,
Luo2,Luo3,Abre0,Fisc0,Abre1,Vali0,Luo4,Fisc1,Fisc3,Mont0,Fisc2,Luo5,Luo6,Gu0,
Gu1,Gu2,Gu3,Fisc4,Luo7,Piki0,Gu4,Gu5,Gupt,Thie0,Mont1,Thie1,Mill1,Fisc5,Gu6,
Gu7,Luo8,Whit,Mill2,Gu8,Gasn}, updated with the latest available information.

\section{Head-on beam--beam compensation}
\begin{figure}[h]
\centering
\includegraphics[width=80mm]{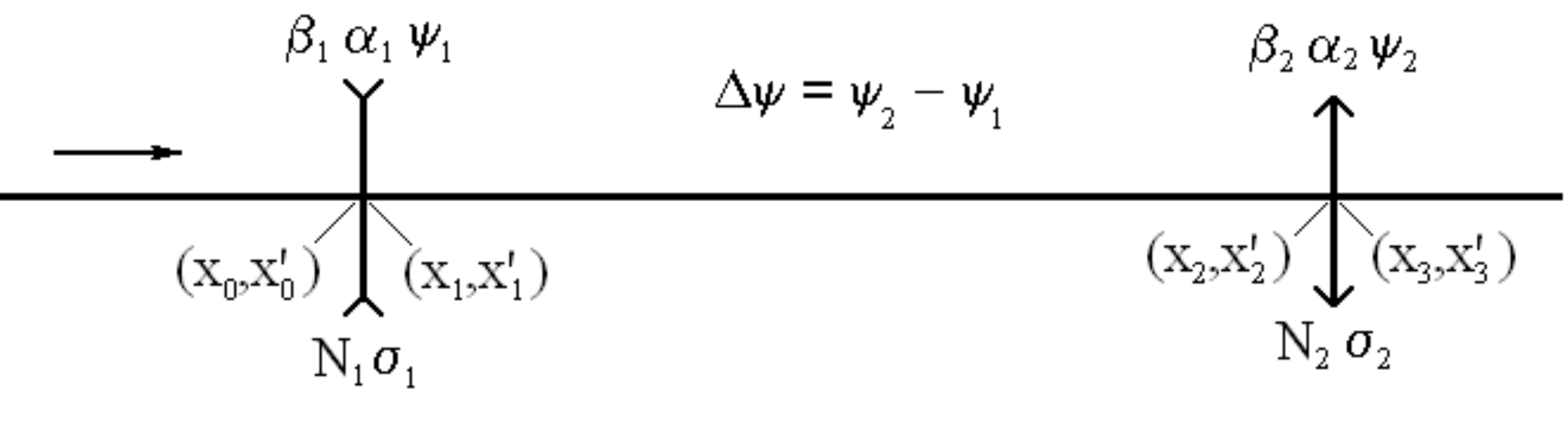}
\caption{Schematic of head-on beam--beam compensation in a beam line view.
At the first location,
with lattice parameters $(\beta_1,\alpha_1,\psi_1)$, a proton experiences a
beam--beam kick from another proton bunch with intensity $N_1$ and root-mean-square beam
size $\sigma_1$. At the second location, with lattice parameters
$(\beta_2,\alpha_2,\psi_2)$, another beam--beam kick is generated by the
electron beam with effective bunch intensity $N_2$ and root-mean-square beam size
$\sigma_2$.}
\label{fig:HOBBC_bl}
\end{figure}

\begin{figure}[h]
\centering
\includegraphics[width=80mm]{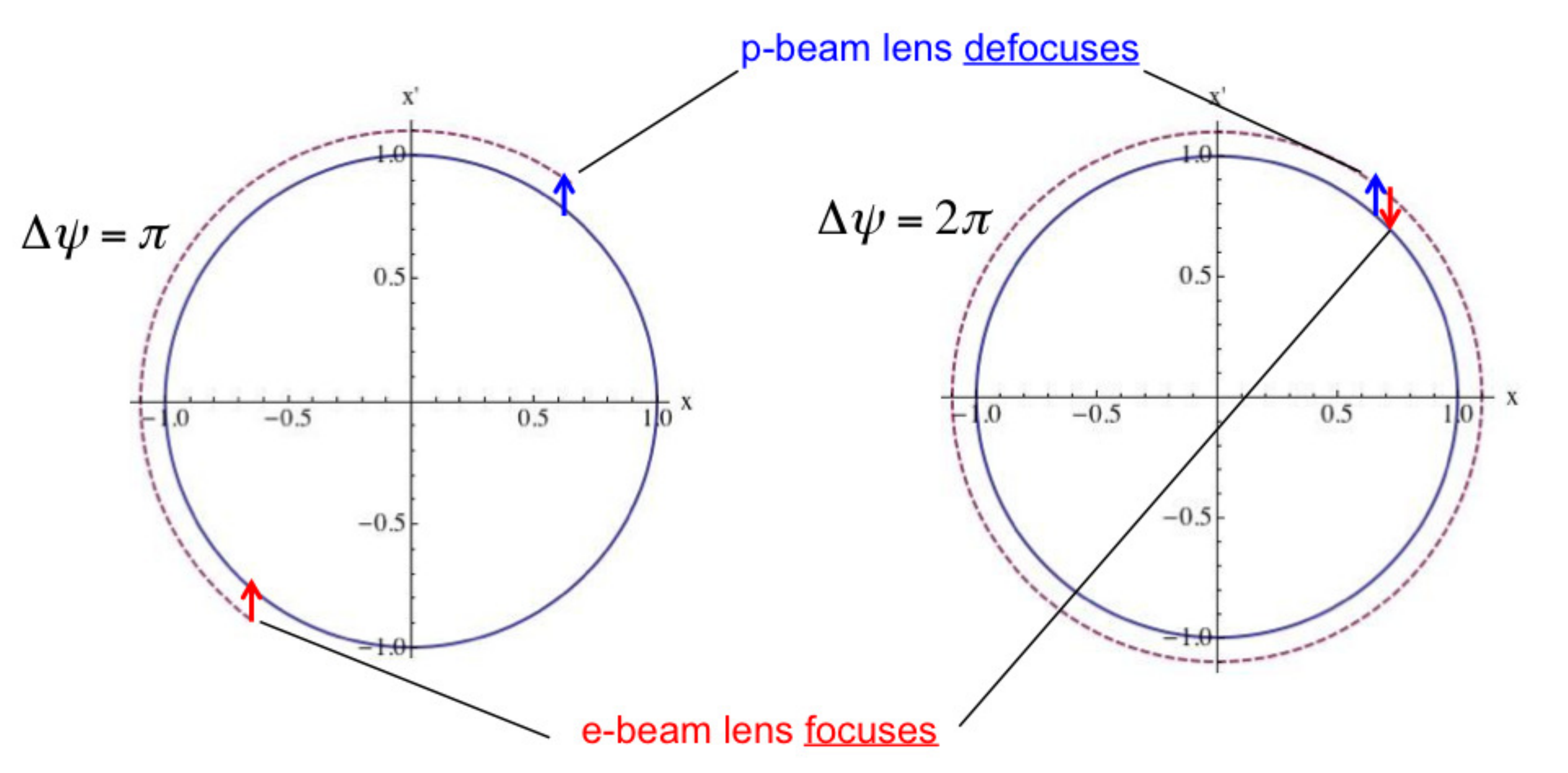}
\caption{Schematic of head-on beam--beam compensation in a normalized phase space view.}
\label{fig:HOBBC_ps}
\end{figure}

If a collision of a proton beam with another proton beam is followed by a
collision with an electron beam, the head-on beam--beam kick can in principle
be reversed. For simplicity we  consider only the horizontal plane and beams
with a Gaussian transverse distribution. Figure~\ref{fig:HOBBC_bl} shows the
beam line layout for head-on compensation, and Fig.~\ref{fig:HOBBC_ps} shows the
normalized phase space view.

Before experiencing a beam--beam kick from another proton beam at location~1,
a proton has transverse phase space coordinates $(x_0,x_0')$. The
proton then receives a kick from the other proton beam~\cite{Keil},
\begin{equation*}
\vspace*{-2mm}
  \Delta x_0' = \frac{2N_1 r_0}{\gamma x_0}
  \left[1-\exp \left(-\frac{x_0^2}{2\sigma_1^2}\right) \right],
\end{equation*}
where $N_1$ is the bunch intensity of the second proton beam, $\gamma$ is the
relativistic factor of the proton receiving the kick, $r_0$ is the classical
proton radius, and $\sigma_1$ is the root-mean-square (rms)
 beam size of the second proton beam. The
new coordinates are then
\begin{align*}
\vspace*{-3mm}
  x_1  &= x_0 , \\
  x_1' &= x_0' + \Delta x_0'.
\end{align*}
After transport through the linear beam line, the coordinates are
\begin{align*}
\vspace*{-3mm}
  x_2  &= M_{11}x_1 + M_{12}x_1',\\
  x_2' &= M_{21}x_1 + M_{22}x_1',
\end{align*}
with (see~\cite{Cour,Syph0})
\begin{align*}
\vspace*{-2mm}
  M_{11} &= \sqrt{\frac{\beta_2}{\beta_1}}
    \, (\cos\Delta\psi + \alpha_1\sin\Delta\psi) ,\\
  M_{12} &= \sqrt{\beta_1\beta_2}\,\sin\Delta\psi ,\\
  M_{21} &= -\frac{1+\alpha_1\alpha_2}{\sqrt{\beta_1\beta_2}}
    \sin\Delta\psi + \frac{\alpha_1-\alpha_2}{\sqrt{\beta_1\beta_2}}
    \cos\Delta\psi ,\\
  M_{22} &= \sqrt{\frac{\beta_1}{\beta_2}}\,(\cos\Delta\psi -
    \alpha_2\sin\Delta\psi)
\end{align*}
where  $\Delta\psi = \psi_2 - \psi_1$. In the electron lens, the proton receives
the kick
\vspace*{-2mm}
\begin{equation*}
\vspace*{-1mm}
  \Delta x'_2 = -\frac{2N_2 r_0}{\gamma x_2}
  \left[1-\exp \left(-\frac{x^2_2}{2\sigma_2^2}\right) \right],
\end{equation*}
where $N_2$ is the effective bunch intensity of the electron lens beam
(i.e.\ the number of electrons the proton passes in the lens) and
$\sigma_2$ is the rms\ beam size of the electron lens beam. The coordinates
after passing the electron lens are then
\begin{align*}
\vspace*{-2mm}
  x_3  &= x_2 ,\\
  x_3' &= x_2' + \Delta x_2'.
\end{align*}
One can now express the final coordinates $(x_3,x_3')$ as a function of the
intensities $(N_1,N_2)$ and require, for exact compensation, that
\begin{equation}
\vspace*{-2mm}
  x_3(N_1,N_2)  = x_3(0,0)   \label{eq:c1}
  \end{equation}
  and
  \begin{equation}
  x_3'(N_1,N_2) = x_3'(0,0),   \label{eq:c2}
\end{equation}
i.e.\ that the final coordinates are the same with and without beam--beam interaction
and compensation.
From the condition~(\ref{eq:c1}) it follows that $M_{12}=0$ and hence
$\Delta\psi = k\cdot\pi$, where $k$ is  an integer. From the
condition~(\ref{eq:c2}) it follows that $N_1=N_2$ and $\sigma_1^2/\sigma_2^2
= \beta_1/\beta_2$.

Therefore, if the following three conditions are met, the beam--beam kicks are
cancelled exactly.
\begin{enumerate}
\vspace*{-2mm}
\item The ion beam and electron beam produce the same amplitude-dependent
force by having the same effective charge and profile.
\vspace*{-2mm}
\item The phase advance between the two beam--beam collisions is a multiple
of $\pi$ in both transverse planes.
\vspace*{-2mm}
\item There are no nonlinearities between the two collisions.
\end{enumerate}
In practice the above can be achieved only approximately.

\noindent Deviations from condition 1 include:
\begin{itemize}
  \vspace*{-2mm}
  \item an electron current that does not match the proton bunch intensity;
  \vspace*{-2mm}
  \item a non-Gaussian electron beam profile (assuming that the proton beam
    transverse profile is Gaussian);
  \vspace*{-2mm}
  \item an electron beam size that differs from the proton beam size;
  \vspace*{-2mm}
  \item time-dependence of the electron and proton beam parameters.
  \vspace*{-2mm}
\end{itemize}
\noindent Deviations from condition 2 include:
\begin{itemize}
  \vspace*{-2mm}
  \item a phase advance $\Delta\psi \ne k\pi$ between the head-on collision
    and the electron lens;
  \vspace*{-2mm}
  \item long bunches, i.e.\ $\sigma_{\rm s} \gtrapprox \beta^*$.
  \vspace*{-2mm}
\end{itemize}
\noindent Deviations from condition 3 include:
\begin{itemize}
  \vspace*{-2mm}
  \item lattice sextupoles and octupoles, as well as multipole error between
    the head-on collision and the electron lens.
  \vspace*{-2mm}
\end{itemize}
Tolerances were studied extensively in simulations and reported in
Ref.~\cite{Luo8}, and bunch length effects have been investigated in
Refs.~\cite{Fisc1,Fisc3}.
The Tevatron experience also provides tolerances for positioning errors,
transverse shape and size mismatches, and electron current variations.
We give the tolerances for all devices below.

We plan to compensate for only one of the two head-on collisions in RHIC, since
a full compensation would lead to a small tune spread and could
give rise to instabilities.

\section{RHIC electron lens design}
In designing the electron lens, we were aiming for a technically feasible
implementation that would come as close as possible to the ideal compensation
scheme outlined above. In addition, a major design consideration was ease
of commissioning and operation. Our goal is a commissioning that is largely parasitic
to the RHIC operation for physics. The main design process can be summarized
as follows.

\begin{table}[tbh]
\begin{center}
\caption{Reference cases for RHIC beam--beam and beam-lens interactions.
Bunch intensities without electron lenses are expected to saturate at
about $2\times 10^{11}$ because of head-on beam--beam effects~\cite{Luo8,RHICbb6}.}
\label{tab:ref}
\vspace*{2mm}
\footnotesize
\begin{tabular}{lcccc}
\hline\hline
Quantity & Unit & \multicolumn{3}{c}{Value} \\
\hline
{\bf Proton beam parameters}\\
Total energy $E_{\rm p}$                   & GeV       & 100 & 255 & 255 \\
Bunch intensity $N_{\rm p}$                & $10^{11}$  & 2.5  & 2.5  & 3.0 \\
$\beta^*_{x,y}$ at IP6, IP8 (p--p)    & m          & 0.85 & 0.5  & 0.5 \\
$\beta^*_{x,y}$ at IP10 (p--e)        & m          & 10.0 & 10.0 & 10.0 \\
Lattice tunes $(Q_x,Q_y)$     & -- & \multicolumn{3}{c}{--- (0.695,\,0.685) ---} \\
rms emittance $\epsilon_n$, initial  & \hspace*{-2mm}mm$\,$mrad & \multicolumn{3}{c}{--- 2.5 ---} \\
rms beam size at IP6, IP8 $\sigma^*_{\rm p}$ & $\mu$m   & 140   & 70    & 70    \\
rms beam size at IP10     $\sigma^*_{\rm p}$ & $\mu$m   & 485   & 310   & 310   \\
rms bunch length $\sigma_{\rm s}$          & m          & 0.50  & 0.40  & 0.20  \\
Hourglass factor $F$, initial        & --        & 0.88  & 0.85  & 0.93  \\
Beam--beam parameter $\xi$/IP         & --       & 0.012 & 0.012 & 0.015 \\
Number of beam--beam IPs      & -- & \multicolumn{3}{c}{--- 2\,+\,1$^{\rm a}$ ---}\\
\hline
{\bf Electron lens parameters}\\
Distance of centre from IP           & m   & \multicolumn{3}{c}{--- 2.0 ---} \\
Effective length $L_{\rm e}$               & m   & \multicolumn{3}{c}{--- 2.1 ---} \\
Kinetic energy $E_{\rm e}$                 & keV        & 7.8  & 7.8  & 9.3 \\
Relativistic factor $\beta_{\rm e}$        & --        & 0.18 & 0.18 & 0.19 \\
Electron line density $n_{\rm e}$ & \hspace*{-2mm}$10^{11}\,$m$^{-1}$ & 1.0  & 1.0  & 1.2 \\
Electrons in lens $N_{\rm e1}$           & $10^{11}$   & 2.1  & 2.1  & 2.5 \\
Electrons encountered $N_{\rm e2}$ & $10^{11}$   & 2.5  & 2.5  & 3.0 \\
Current $I_{\rm e}$                        & A          & 0.85 & 0.85 & 1.10 \\
\hline\hline
\multicolumn{5}{l}{$^{\rm a}$\,{\footnotesize One head-on collision in IP6 and IP8 each,
plus a compensating}}\\
\multicolumn{5}{l}{\footnotesize  head-on collision in IP10.}
\end{tabular}
\end{center}
\end{table}

\textbf{Condition 1} (same amplitude-dependent forces
from the proton beam and electron lens) has a number of implications. Since both
proton beams are round in the beam--beam interactions ($\beta^*_x=
\beta^*_y$ and $\epsilon_x=\epsilon_y=\epsilon_n$), we also require that  $\beta_x=
\beta_y$ at the electron lens location, and require matched transverse proton and
electron beam profiles, i.e.\ that the electron beam profile is also Gaussian with
$\sigma_{{\rm p},x}=\sigma_{{\rm e},x}=\sigma$ and $\sigma_{{\rm p},y}=\sigma_{{\rm e},y}=\sigma$.
The condition $\beta_x=\beta_y$ limits the electron
lens locations to the space between the DX magnets; in these locations the
RHIC lattice also has a small dispersion.

The tolerances for the main solenoid field straightness and for the relative
beam alignment are easier to meet with a larger proton beam. A larger beam is
also less susceptible to coherent instabilities~\cite{TEL2,Whit}.
The $\beta$-function at IP10 cannot be larger than 10~m at 250~GeV
proton energy without modifications to the  buses
and feedthroughs of the IR10 superconducting magnets. Such modifications
are currently not under consideration  because of costs, but could be implemented if
coherent instabilities occur and cannot be mitigated by other means.

With a fully magnetized electron beam, the beam size $\sigma_{\rm e}$  in the main solenoid
is given by its size at the cathode, $\sigma_{\rm ec}$, together with the solenoid
fields $B_{\rm sc}$ at the cathode  and $B_{\rm s}$ in the main solenoid  as
$\sigma_{\rm e} = \sigma_{\rm ec}\sqrt{B_{\rm sc}/B_{\rm s}}$. For technological and cost reasons,
the field $B_{\rm s}$ cannot be much larger than 6~T, and a strong field makes a
correction of the field straightness more difficult. The field $B_{\rm sc}$ has to
be large enough to suppress space charge effects. With the limits in the
$B_{\rm sc}$ and $B_{\rm s}$ fields and a given beam size $\sigma_{\rm e}$, the electron
beam size and current density at the cathode follow, and they must be technically
feasible. Unlike the Tevatron electron lenses, we use a DC electron beam to
avoid the noise possibly introduced through the high-voltage switches.
A DC beam requires the removal of ions created in the electron lens through
residual gas ionization.

\textbf{Condition 2} (phase advance of multiples of $\pi$ between p--p and p--e
interaction) can be achieved through lattice modifications. We have installed
four phase-shifter power supplies for both transverse planes of both rings
so that the betatron phase between IP8
and the electron lenses in IR10 can be adjusted. To have $\Delta \psi=k\pi$
in both planes of both rings, it is  also necessary to change the integer tunes
from $(28,29)$ to $(27,29)$ in the Blue ring and from $(28,29)$ to $(29,30)$
in the Yellow ring  to find a solution. With the new lattices, higher
luminosities were reached in 2013 than in previous years, but
the polarization was lower. The lower polarization is still being investigated and
may not have resulted from the new lattices. Other lattice options are
also under study: (i) a solution was found for the Yellow ring that maintains
the integer tunes of $(28,29)$ and has the correct phase advances; (ii) the
phase advance of a multiple of $\pi$ may also be realized between IP6 and
the electron lenses.

{\bf Condition 3} (no nonlinearities between the p--p and p--e interactions)
is best achieved when the p--e interaction is as close as possible to the p--p
interaction. With the location in IR10 (Fig.~\ref{fig:ov}), there is only one
arc between the p--p interaction at IP8 and the p--e interaction at IR10. In
this configuration, a proton, after receiving a beam--beam kick at IP8, passes
a triplet with nonlinear magnetic fields from field errors, an arc with
chromaticity sextupoles and dodecapoles in the quadrupoles as dominating
nonlinear field errors, and another triplet in IR10.
To avoid bunch length effects, the parameter $\beta^*$ cannot be too small~\cite{Fisc1,Fisc3}.
In simulations, a value as low as $\beta^*=0.5$~m was found to be
acceptable~\cite{Luo8}.

\begin{figure}[h]
\centering
\includegraphics[width=80mm]{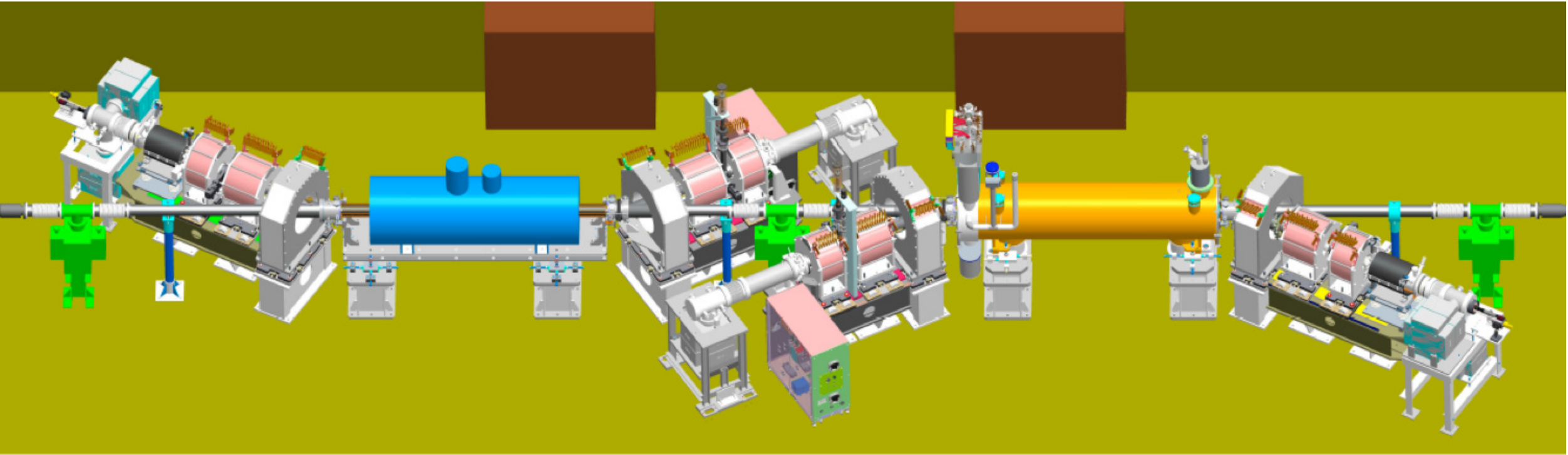}
\caption{Layout of the two electron lenses in IR10. In 2013 the Blue
lens (left) had the EBIS spare solenoid installed instead of the
superconducting solenoid designed for the electron lens. In each lens
three beams are present, the two proton beams and the electron beam acting
on one of the proton beams; the proton beams are vertically separated. }
\label{fig:3}
\end{figure}

\smallskip
The location of both the Blue and the Yellow electron lenses in IR10, in a section
common to both beams (Fig.~\ref{fig:3}), allows local compensation of
the main solenoid effect on both linear coupling and spin orientation by having
the two main solenoids with opposing field orientations. At 255~GeV proton
energy, one superconducting solenoid with a 6~T field introduces coupling
that leads to $\Delta Q_{\min} = 0.0023$ \cite{Luo5} and increases all spin
resonance strengths by 0.003~\cite{Bai}. In this configuration it is also
possible to ramp the magnets together during RHIC stores without affecting the
beam lifetime or spin orientation.

The instrumentation must allow for monitoring of the electron beam current and
shape as well as the relative position and angle of the electron and proton
beams in the electron lens. Two modes are foreseen: a setup mode in which the
electron beam current is modulated and affects only a single bunch in RHIC,
and a compensation mode with a DC electron beam. The main parameters of the
electron beams are presented in Table~\ref{tab:ref}. 

\begin{figure*}[tbh]
\centering
\includegraphics[width=155mm]{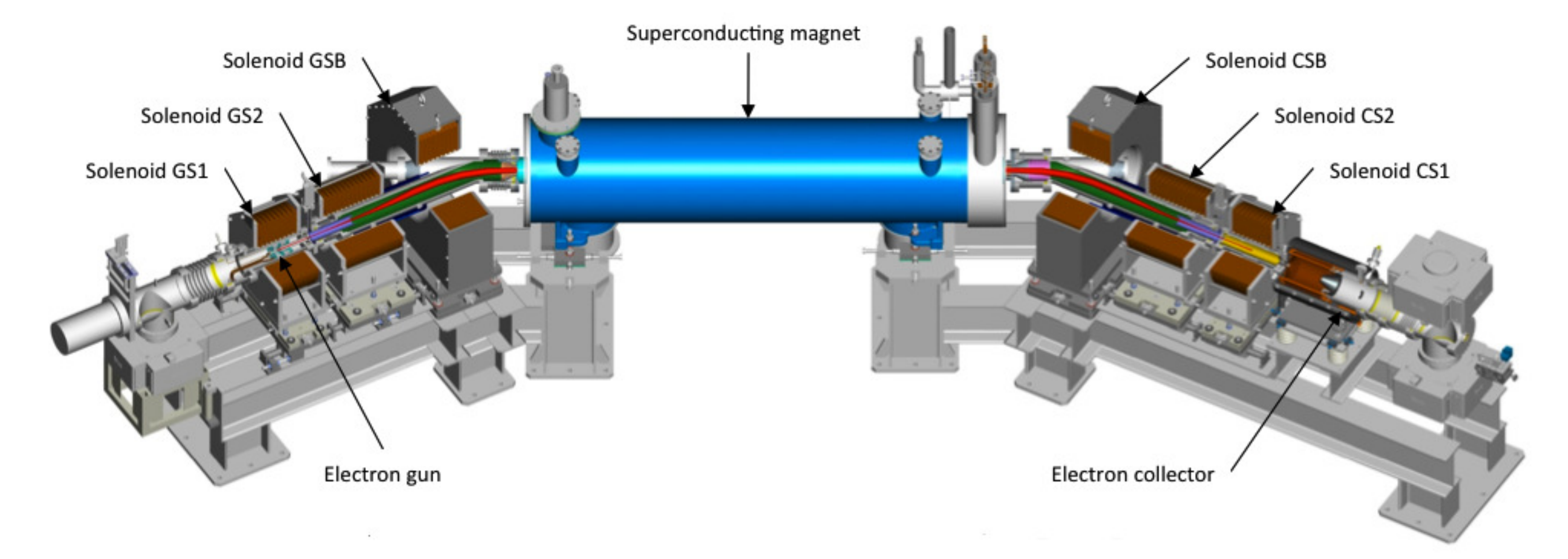}
\caption{RHIC electron lens. The electrons in the DC beam move from left to
right and interact with the protons, which move in the opposite direction,
inside the superconducting solenoid.}
\label{fig:elens}
\end{figure*}

A RHIC electron lens consists of (see Fig.~\ref{fig:elens}) an electron gun, an
electron beam transport to the main solenoid, the superconducting main
solenoid in which the interaction with the hadron beam occurs, an electron
beam transport to the collector, an electron collector, and instrumentation.

\subsection{Electron Gun}
\begin{figure}[htb]
\centering
\includegraphics[width=68mm]{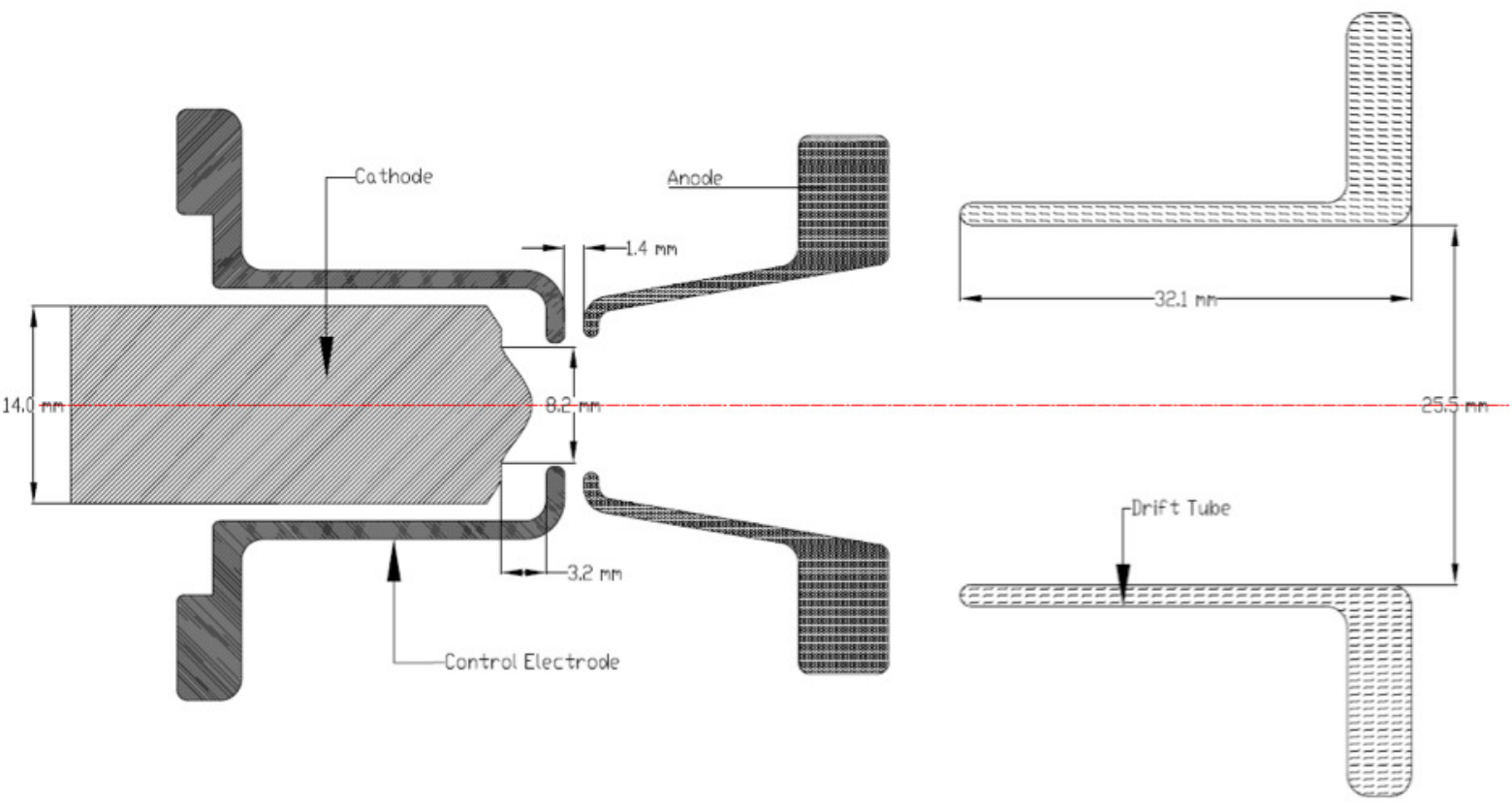}
\includegraphics[width=68mm]{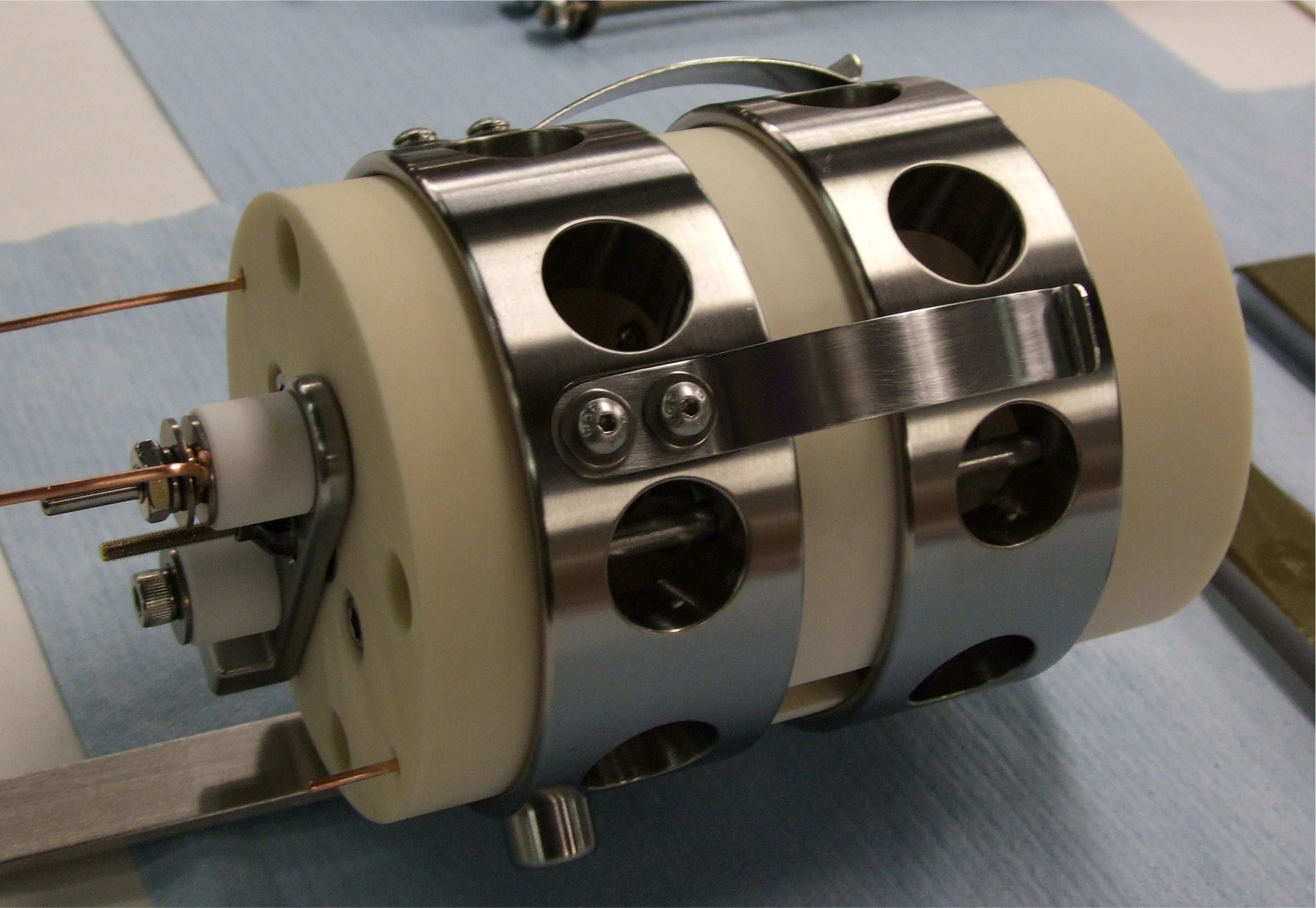}
\caption{Gun schematic (top) and manufactured gun (bottom).}
\label{fig:egun}
\end{figure}

The electron gun (see Fig.~\ref{fig:egun} and Table~\ref{tab:egun})~\cite{Piki0} has to
provide a beam with a transverse profile that is close to Gaussian. Considering
the magnetic compression of the electron beam into the main solenoid centre
with a maximum magnetic field of 6.0~T, a cathode radius of 4.1~mm gives a
Gaussian profile with 2.8 rms beam sizes. The perveance of the gun is
$P_{\rm gun}=1.0\times 10^{-6}$~AV$^{-1.5}$. The current density of the electron beam
on its radial periphery can be changed with the control electrode voltage
(Fig.~\ref{fig:egun}, top), while the general shape of the beam profile remains
Gaussian. The cathodes (LB$_6$ and IrCe) were produced at BINP in
Novosibirsk~\cite{Kuzn}.
With a nominal current density of 12~A/cm$^2$,  IrCe was chosen as the cathode
material for its long lifetime (greater than 10\,000~h).

An assembled gun is shown in the bottom panel of Fig.~\ref{fig:egun}. The gun has three
 operating  modes: (i)~DC for continuous compensation; (ii)~100~Hz for electron beam
positioning with BPMs, such that the electron current rises between the last two RHIC
bunches and falls in the abort gap; (iii)~78~kHz for single-bunch compensation, with rise and fall time as in the 100~Hz mode.

The gun and collector vacuum is UHV compatible, with a design pressure
of 10$^{-10}$ Torr and  a nominal
pressure of 10$^{-11}$ Torr for the interface to the RHIC warm bore.
For this reason, all of the components are
bakeable to 250$^\circ$C. The gun and collector chambers will have a confined
gas load by using a conductance-limiting aperture and enough installed
pumping speed. All vacuum chambers interfacing with the RHIC warm bore will
be made from stainless steel.

\begin{table}[tbh]
\begin{center}
\caption{Main parameters of the thermionic electron gun.}
\label{tab:egun}
\vspace*{2mm}
\begin{tabular}{lcc}
\hline\hline
Quantity       & Unit & Value \\
\hline
Perveance      & $\mu$A\,V$^{-3/2}$ & 1.0 \\
Voltage        & kV              & 10  \\
Current        & A               & 1.0 \\
Profile        & --             & Gaussian  \\
Cathode radius & mm\,/\,$\sigma$   & 4.1\,/\,2.8 \\
Max B-field    & T               & 0.8 \\
Modes          & --           & DC, 100~Hz, 78~kHz \\
\hline\hline
\end{tabular}
\end{center}
\end{table}

\subsection{Electron Collector}
\begin{figure}[htb]
\centering
\includegraphics[width=70mm]{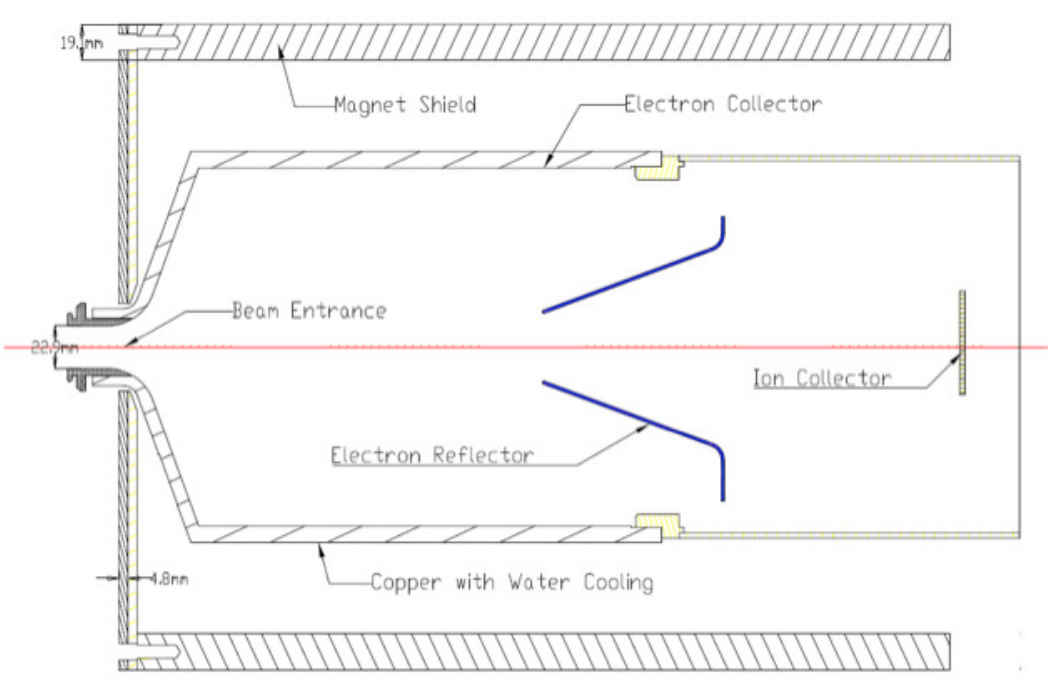}
\includegraphics[width=70mm]{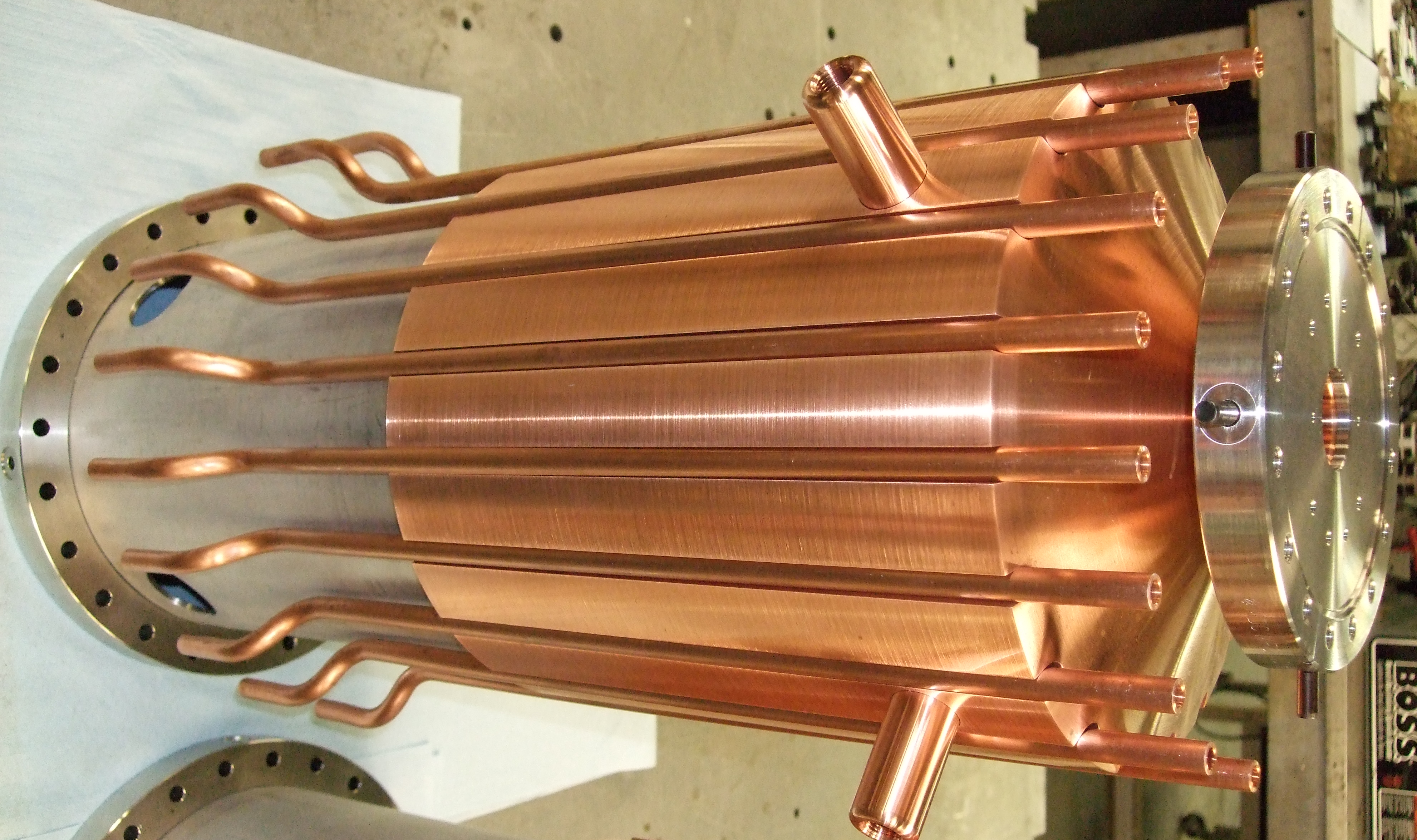}
\caption{Collector schematic (top) and collector during manufacturing
(bottom).}
\label{fig:ecoll}
\end{figure}

The collector spreads the electrons on the inside of a cylindrical surface
that is water-cooled on the outside (see Fig.~\ref{fig:ecoll}). Simulations
give a power density of 10~W/cm$^2$ for a 10~A electron beam, decelerated to
4~keV. The collector can absorb up to four times this power density~\cite{Piki0}.
The design is
dictated primarily by the UHV requirements of RHIC. It separates
the heavily bombarded area from the rest of the electron lens by using a small
diaphragm. A magnetic shield leads to fast diverting electrons inside the
collector. The reflector has a potential lower than the cathode and pushes
electrons outwards to the water-cooled cylindrical surface. Under a load twice
as high as expected from a 2~A electron beam, the maximum temperature on the inner
surface of the shell is 102$^\circ$C. This temperature is acceptable for the
material (copper) and for UHV conditions in RHIC. Twenty tubes with an ID~=~8.0~mm
are brazed to the outside of the cylindrical shell and are connected in
parallel for water flow (Fig.~\ref{fig:ecoll}).

The collector design also limits the flow of secondary and backscattered
electrons from the collector towards the interaction region because the volume
is magnetically shielded.

The gun and collector power supplies are referenced to the cathode. The gun
supplies include the cathode bias supply, the cathode heater, the beam-forming
supply, and two anode supplies (DC and pulsed). The collector power supply is
rated with 10~kV at 2~A, and will limit the energy deposited in the device
should an arc occur. An ion reflector is powered with respect to the cathode
potential. A suppressor element is powered with respect to the collector.

\subsection{Superconducting Main Solenoid}
\begin{figure}[tbh]
\centering
\includegraphics[width=82mm]{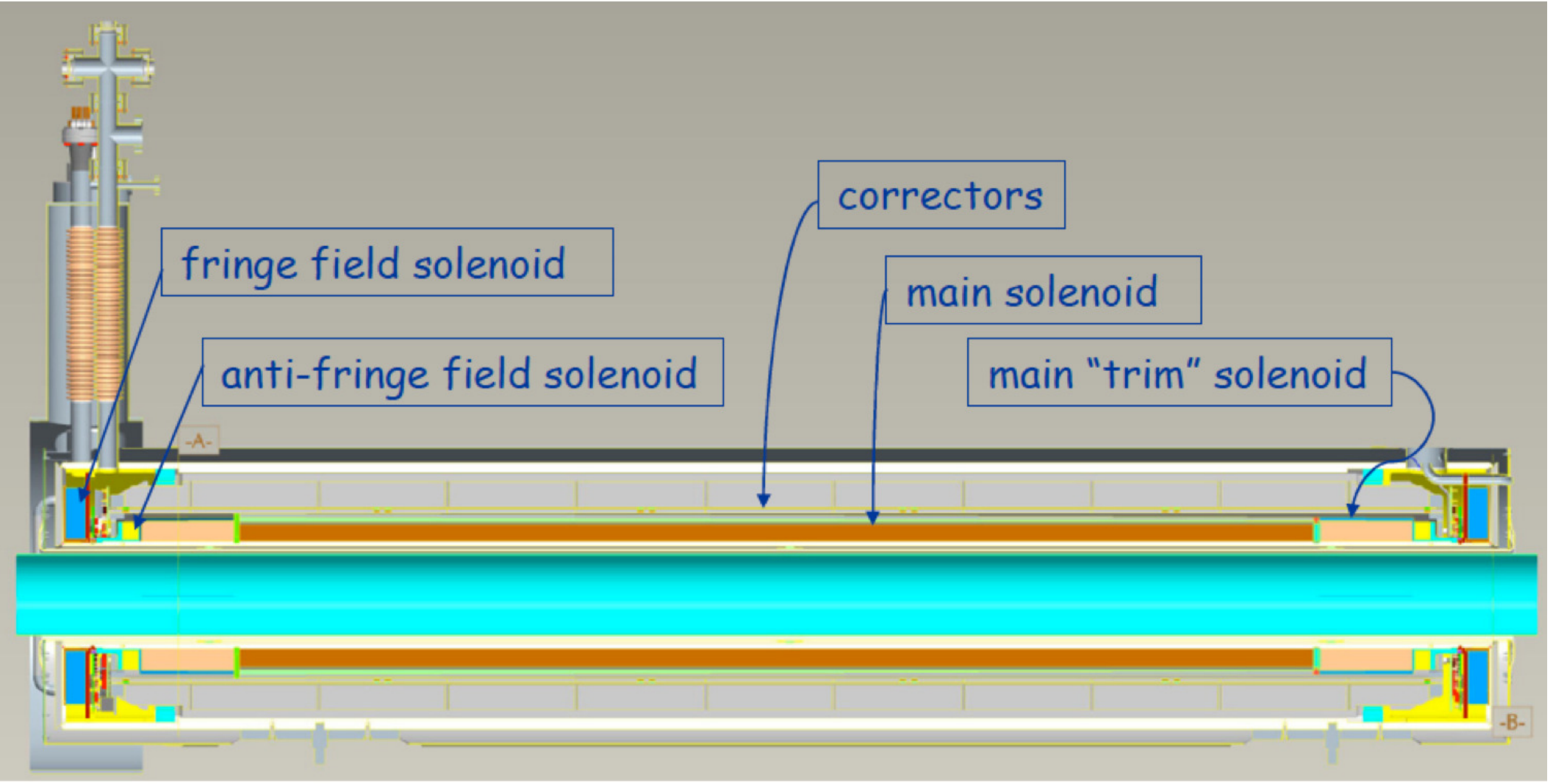}
\caption{Superconducting main solenoid with fringe and anti-fringe solenoids,
and straightness and angle correctors.}
\label{fig:scs}
\end{figure}

A superconducting solenoid guides and stabilizes the low-energy electron
beam during the interaction with the proton beam, and allows for magnetic
compression of the electron beam size to the proton beam size. The
superconducting main solenoid is a warm bore magnet with an operating
field of 1--6~T (Fig.~\ref{fig:scs}). The cryostat includes a number of
additional magnets for a total of 17~\cite{Gupt}.
The main parameters are given in Table~\ref{tab:sms}.

\begin{table}[tbh]
\vspace*{-2mm}
\begin{center}
\caption{Main parameters of the superconducting solenoids and
corrector magnets in the same cryostat.}
\label{tab:sms}
\vspace*{2mm}
\footnotesize
\begin{tabular}{lcc}
\hline\hline
Quantity & Unit & Value \\
\hline
Cryostat length                   & mm     & 2838        \\
Coil length                       & mm     & 2360        \\
Warm bore inner diameter          & mm     & 154         \\
Uniform field region              & mm     & $\pm 1050$  \\
Main coil layers                  & --    & 22 (11 double) \\
Additional trim layers in ends    & --    & 4 (2 double)   \\
Wire $I_{\rm c}$ specification (4.2~K, 7~T) & A  & ${>}\,700$      \\
Operating main field $B_{\rm s}$        & T      & 1--6         \\
Field uniformity $\Delta B_{\rm s}/B_{\rm s}$ & -- & $\pm 0.006$ (1--6~T) \\
Field straightness, after correction & $\mu$m & $\pm 50$ (1--6~T) \\
Straightness correctors (5H\,+\,5V) &T\,m &$\pm0.010$\\
Angle correctors (1H\,+\,1V)        & T\,m &$\pm0.015$\\
Inductance                        & H      & 14          \\
Stored energy (6~T)               & MJ     & 1.4         \\
Current (6~T)                     & A      & 430 (473$^{\rm a}$)  \\
\hline\hline
\multicolumn{3}{l}{
$^{\rm a}\,${\footnotesize First double layer disabled.}}
\end{tabular}
\end{center}
\end{table}

Fringe field (FF) solenoid coils at both ends are included to allow for a
guiding and focusing solenoid field for the electrons of no less than 0.3~T
between the superconducting magnet and the warm transport solenoids GSB and
CSB (see Fig.~\ref{fig:elens}). To achieve the desired field uniformity over a
range of field strengths $B_{\rm s}$, anti-fringe field (AFF) coils are placed next
to the FF coils. The FF and AFF coils on both ends can be powered
independently to avoid forming a magnetic bottle with a low main field $B_{\rm s}$,
which would trap backscattered electrons. Extraction of scattered electrons is
also possible by using a split electrode~\cite{Gu7}.

Included in the cryostat are five short (0.5~m) dipole correctors in both
the horizontal and the vertical planes, to correct the solenoid field
straightness to $\pm 50$~$\mu$m. A long (2.5~m) dipole corrector in each
transverse plane allows  the angle of the electron beam inside
the main magnet to be changed by $\pm 1$~mrad (at 6~T) to align the electron and proton
beams.

To reduce the number of layers in the main, FF, and AFF coils, and thereby
the manufacturing time, a relatively large conductor was chosen, and the
current in these coils was 430~A, 470~A, and 330~A, respectively~\cite{Gupt}.
A total of 17 individual coils (main coil, two FF coils, two AFF coils, ten straightness dipole
correctors, and two angle dipole correctors) can be powered.

The magnet is bath-cooled at a temperature just above 4.5~K, dictated by the
operating pressure of RHIC cryogenic system's main warm return header. The current
leads are all conventional vapour-cooled leads with individual flow controllers.
The magnet's thermal shield and supports intercepts are cooled by the balance
of the boil-off vapour not used by the current leads, which also returns to the
main warm return header. The total flow rate draw from the RHIC cryogenic system is
1.6~g/s for each solenoid. Liquid helium can be supplied from a local Dewar
when the RHIC refrigerator is not running.

Both magnets were tested vertically and reached 6.6~T, 10\% above the maximum
operating field, after a few training quenches. The magnets are now fully
cryostatted. During the vertical test of the first magnet, a short in the first
layer was detected, and the first double layer was grounded permanently. This
required raising the operating current from 440~A to 473~A.

The field measurement system is under development.
With proton rms beam sizes as small as 310~$\mu$m in the electron lenses, a
deviation by no more than 50~$\mu$m of the solenoid field lines from straight lines is targeted.
A needle-and-mirror system has been constructed that can
be used in the RHIC tunnel to both measure the straightness of the field lines
and verify the correction with the integrated short dipole correctors. The
needle-and-mirror measurement system is being cross-checked with a vibrating
wire system \cite{Jain1} using the second superconducting solenoid.

\subsection{Warm Magnets}
The electron beam is transported from the gun to the main solenoid and from
the main solenoid to the collector through three warm solenoids each
(Fig.~\ref{fig:elens})~\cite{Gu1,Piki0}. These provide focusing with a solenoid
field of at least 0.3~T along the whole transport channel. Within the GS2 and
CS2 solenoids are also horizontal and vertical steering magnets that can move
the beam by $\pm$5~mm in the main solenoid in either plane.

\begin{figure}[htb]
\centering
\includegraphics[width=80mm]{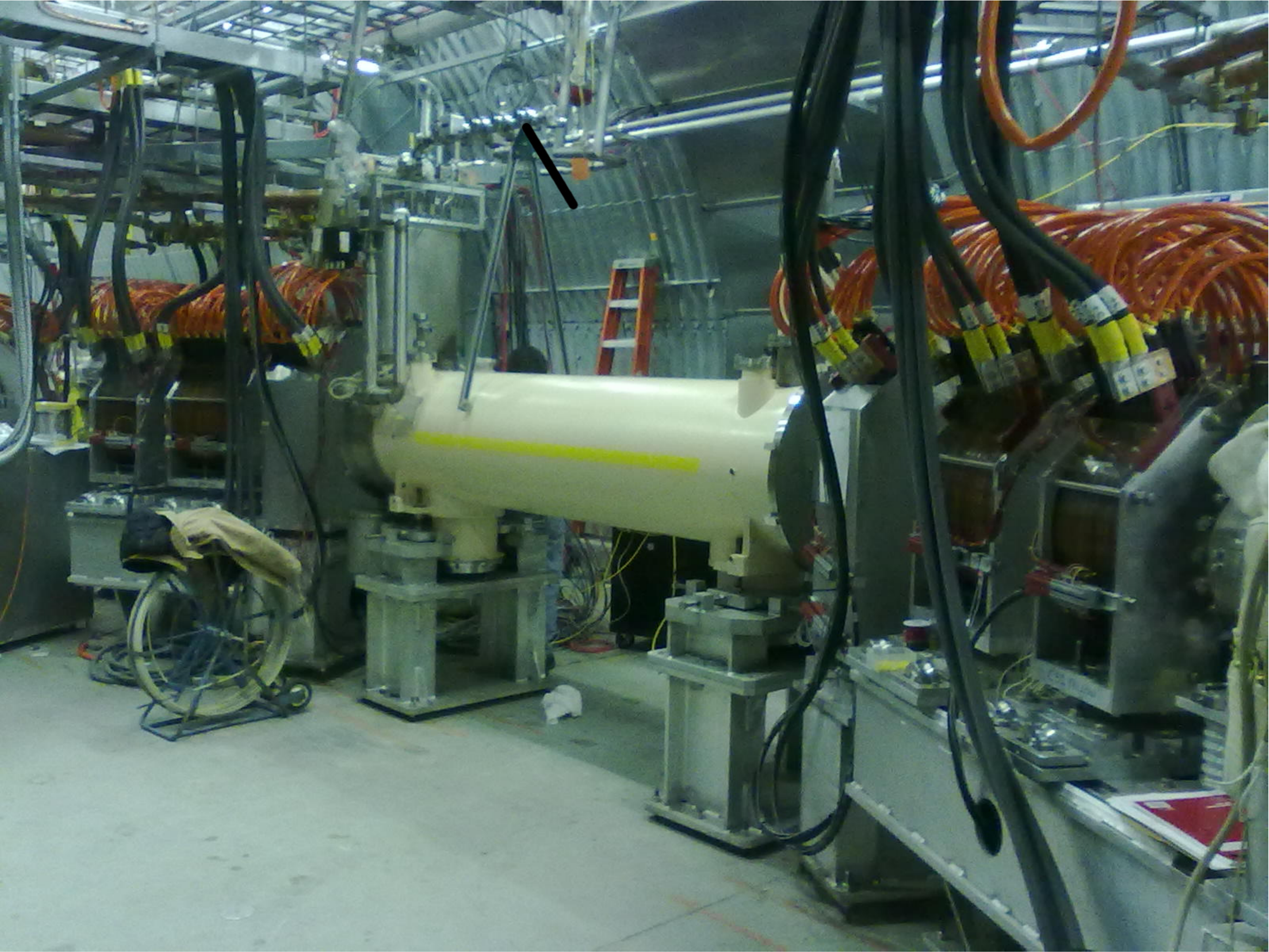}
\caption{Yellow electron lens as installed in 2013. Visible are the gun
side (left), the superconducting main solenoid (centre), and the collector
side (right).}
\label{fig:9}
\end{figure}

The solenoids are made of pancake coils whose field errors have been
optimized~\cite{Gu3}. The power consumption of both electron lenses with
nominal parameters is limited to a total of 500~kW  to avoid upgrades
to the electrical and cooling water infrastructure in IR10. The main parameters
are given in Table~\ref{tab:wm}. All warm magnets and associated power supplies
are installed (Fig.~\ref{fig:9}).

\begin{table}[tbh]
\vspace*{-2mm}
\begin{center}
\caption{Main parameters of the warm magnets.}
\label{tab:wm}
\vspace*{2mm}
\footnotesize
\begin{tabular}{lcccccc}
\hline\hline
Quantity             & Unit   & GS1  & GS2  & GSB & GSX & GSY \\
                     &        & CS1  & CS2  & CSB & CSX & CSY \\
\hline
ID                   & mm     & 174  & 234  & 480 & 194 & 210 \\
OD                   & mm     & 553  & 526  & 860 & 208 & 224 \\
Length               & mm     & 262  & 379  & 262 & 500 & 500 \\
No.~layers            & --    & 13   & 10   & 13  & 12  & 12  \\
No.~pancakes          & --    & 9    & 13   & 9 \\
Inductance           & mH        & 20 & 20 & 40 & 0.2 & 0.2 \\
Resistance           & m$\Omega$ & 40 & 50 & 80 & 20  & 20  \\
Current              & A      & 1188 & 731  & 769  & 258  & 271  \\
Power                & kW     & 58   & 26   & 45   & 1.4  & 1.7  \\
$\Delta T$            & K      & 13.4 & 3.6  & 14.2 & 5.9  & 6.9  \\
$\Delta p$           & bar    & 1.5  & 1.5  & 1.5  & 1.5  & 1.5  \\
Solenoid field $B_{\rm s}$ & T      & 0.8  & 0.45 & 0.32 &      &      \\
\hline\hline
\end{tabular}
\end{center}
\vspace*{-5mm}
\end{table}

\subsection{Instruments and Vacuum System}
The instrumentation monitors the current and shape of the electron beam,
the electron beam losses, and the overlap of the electron beam with the proton beam.
The following items are included (the quantities given in parentheses are for each lens):

\begin{itemize}
\vspace*{-3mm}
  \item dual-plane beam position monitors (2);
\vspace*{-3mm}
  \item e--p beam overlap monitor based on backscattered electrons
    (1)~\cite{Thie1};
\vspace*{-3mm}
  \item differential current monitor (1);
\vspace*{-3mm}
  \item beam loss monitor drift tubes (8);
\vspace*{-3mm}
  \item collector temperature sensor (1);
\vspace*{-3mm}
  \item profile monitor (YAG screen) (1);
\vspace*{-3mm}
  \item profile monitor (pin-hole) (1);
\vspace*{-3mm}
  \item ion collector (1).
\end{itemize}

\begin{figure*}[htb]
\centering
\includegraphics[width=165mm]{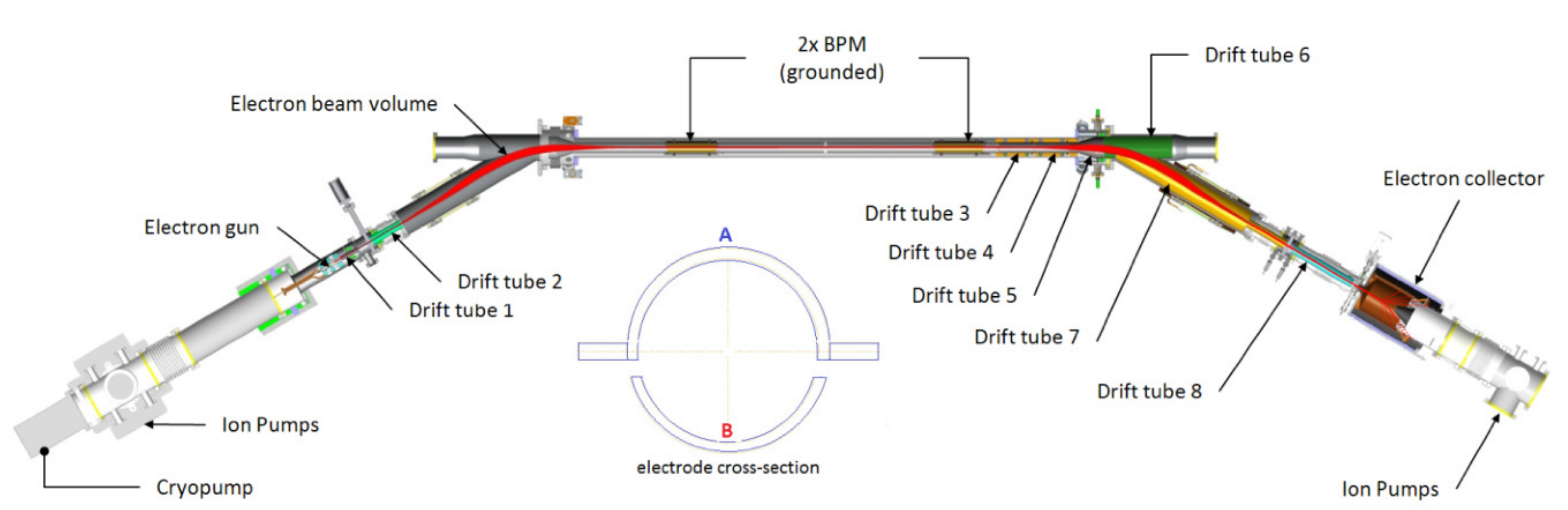}
\caption{Layout of the drift tube system. The inset shows the cross-section of drift tube~4,
which is split for the removal of trapped electrons~\cite{Gu7}. }
\label{fig:dt}
\vspace*{-2mm}
\end{figure*}

\begin{figure}[tbh]
\centering
\includegraphics[width=77mm]{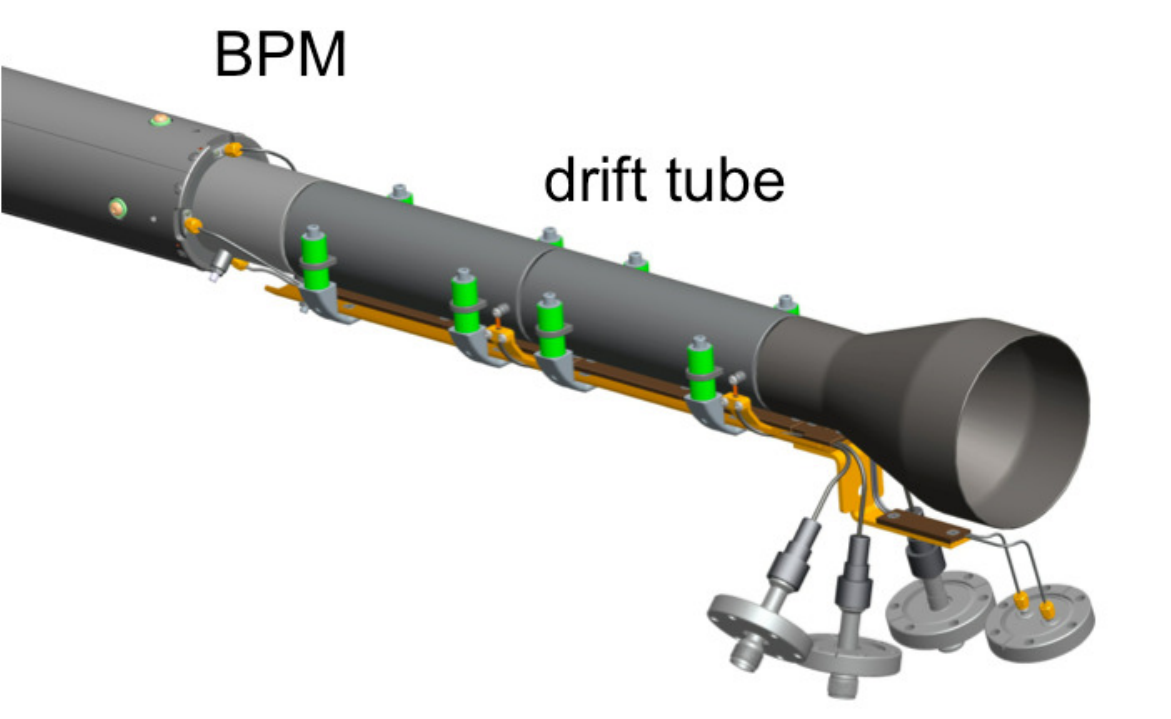}
\caption{Beam position monitor and drift tubes with high-voltage
stand-offs and cable.}
\label{fig:bpm}
\end{figure}

The layout of the vacuum system with the drift tubes is shown in
Fig.~\ref{fig:dt}. A total of eight drift tubes allow for changes in the electron
beam energy and the removal of ions in the interaction region; the split drift
tube~4 enables the removal of backscattered electrons~\cite{Gu7}, which can be
trapped with a low main field $B_{\rm s}$ and high fringe fields. Figure~\ref{fig:bpm}
shows the detail of a section containing  a beam position monitor (BPM), two drift tubes, cables, feedthroughs, and a heat sink
to cool the cables, which can heat up when the proton beam deposits radio-frequency energy in
the structure.

\begin{figure}[tbh]
\centering
\includegraphics[width=80mm]{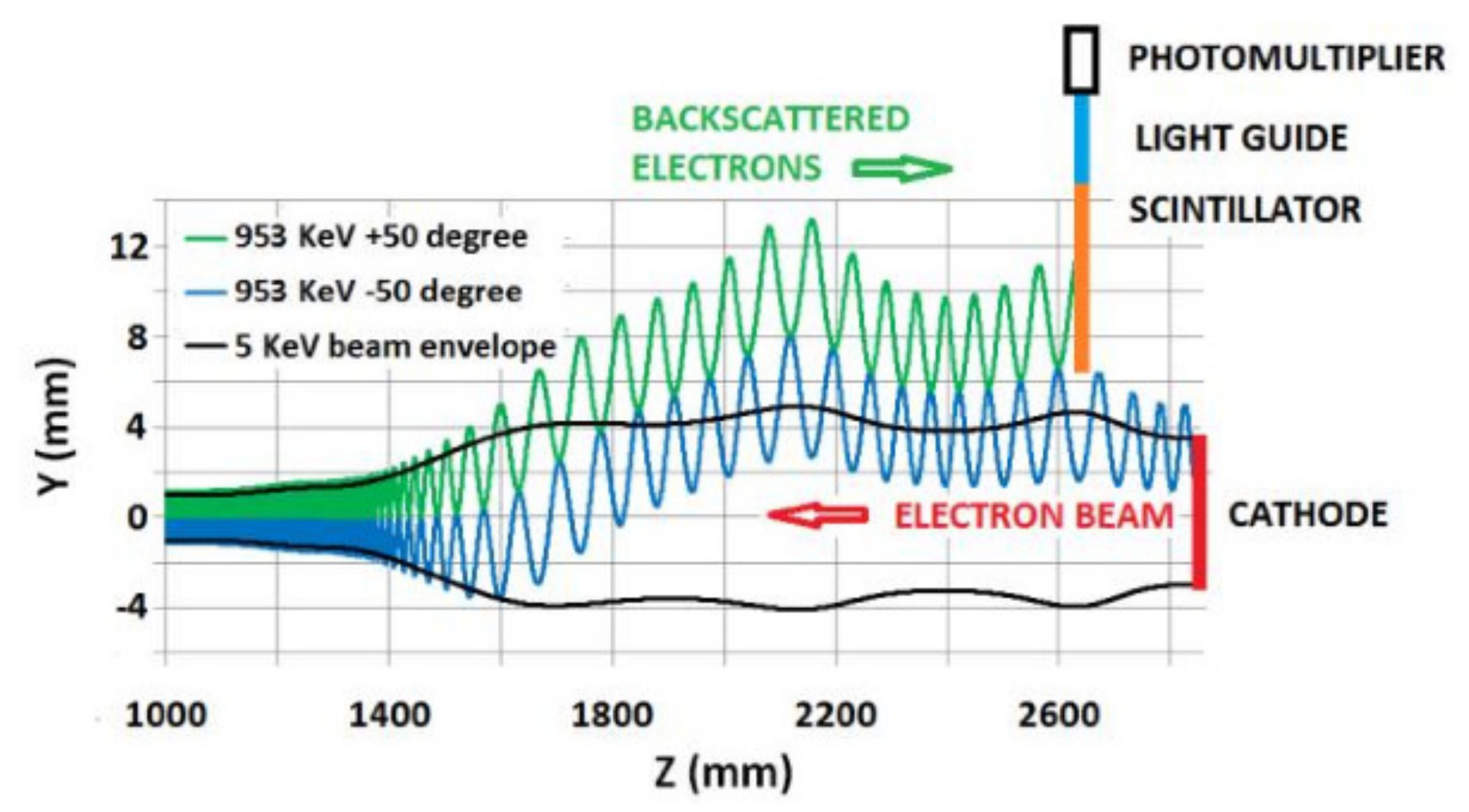}
\includegraphics[width=80mm]{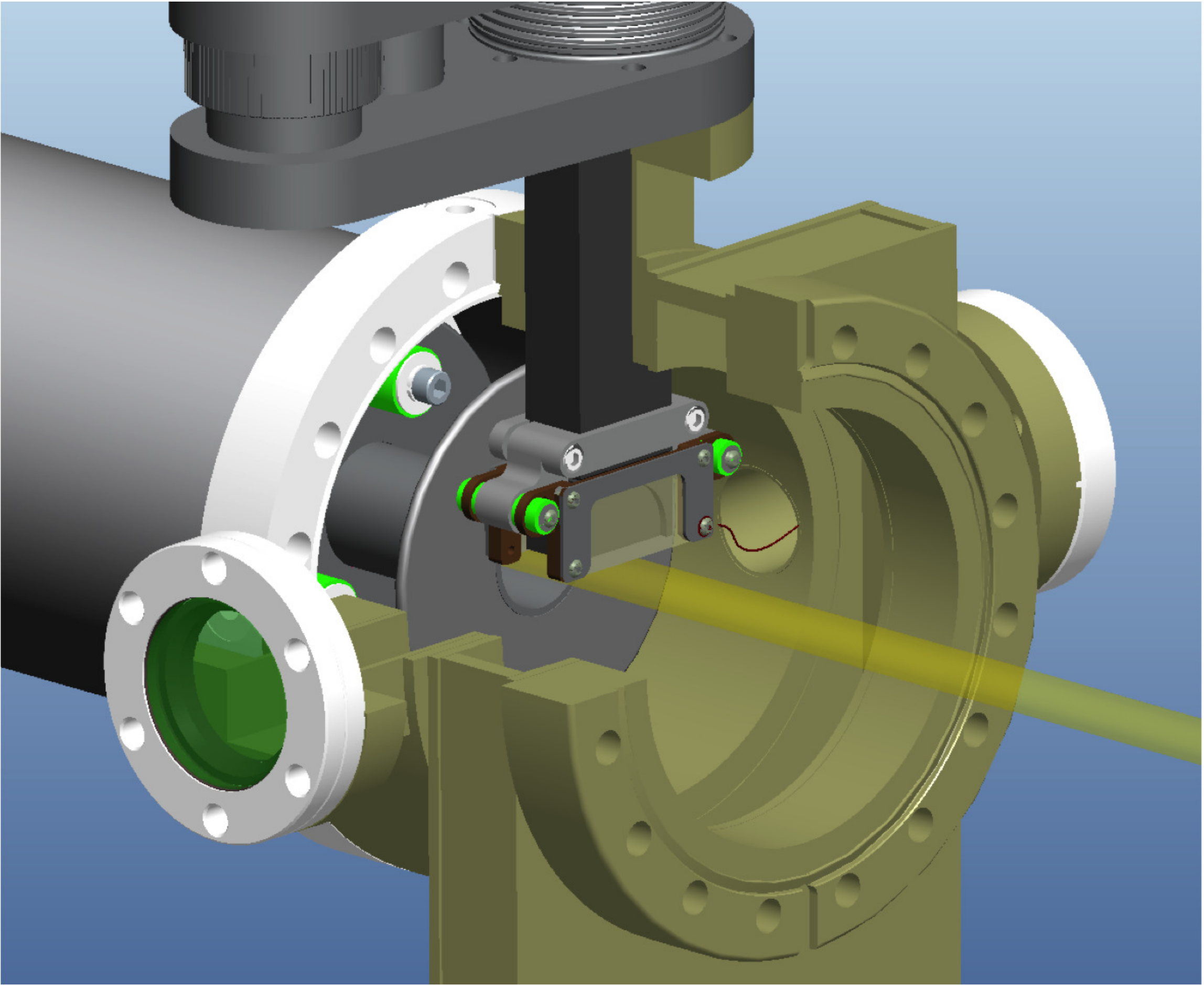}
\caption{Beam overlap monitor using backscattered electrons~\cite{Thie1}.
The top view is a schematic showing two trajectories of backscattered electrons
arriving at the gun above the primary electron beam; the bottom view shows
the positioning mechanism of the detector.}
\label{fig:bs}
\end{figure}

The BPMs see only  a signal with a pulsed beam. The proton beams are bunched,
and a fill pattern can be created so that a bunch in one beam is detected
when there is a gap in the other beam. The electron beam needs to be
pulsed (at 100~Hz or 80~kHz) to be visible. The BPMs are
used to bring the electron and proton beams in close proximity. The final
alignment is done with the beam overlap monitor based on backscattered
electrons~\cite{Thie1}. Alignment was found to be a critical parameter in
the Tevatron electron lenses, and the beams have to be aligned to within a
fraction of the rms beam size, which can be as small as 310~$\mu$m
(see Table~\ref{tab:ref}). Figure~\ref{fig:bs} shows the beam overlap monitor.

\begin{figure}[htb]
\centering
\includegraphics[width=80mm]{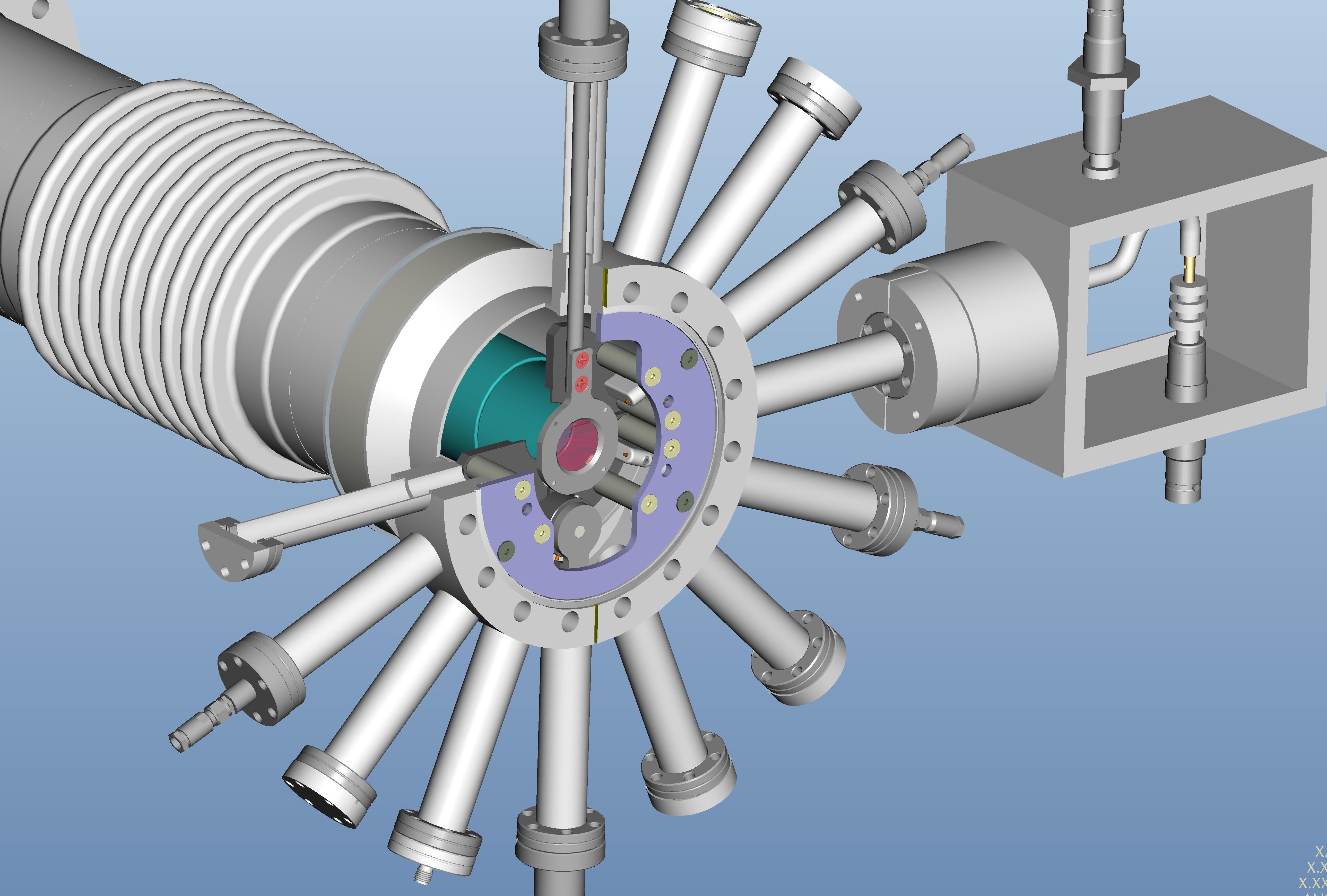}
\caption{Instrument holder in front of the collector. Visible are the
halo detector, YAG screen (inserted), and pin-hole detector (retracted).}
\label{fig:ih}
\vspace*{-2mm}
\end{figure}

The differential current monitor, drift tubes, ion collector, and
collector temperature sensor all monitor the electron beam loss in the lens.
The YAG screen and pin-hole profile monitors can only be used in a low-power mode.
The extracted ion current is monitored in a collector~\cite{Piki0}.

\vspace*{-1.0mm}
\section{Test bench results}
\begin{figure}[htb]
\centering
\includegraphics[width=80mm]{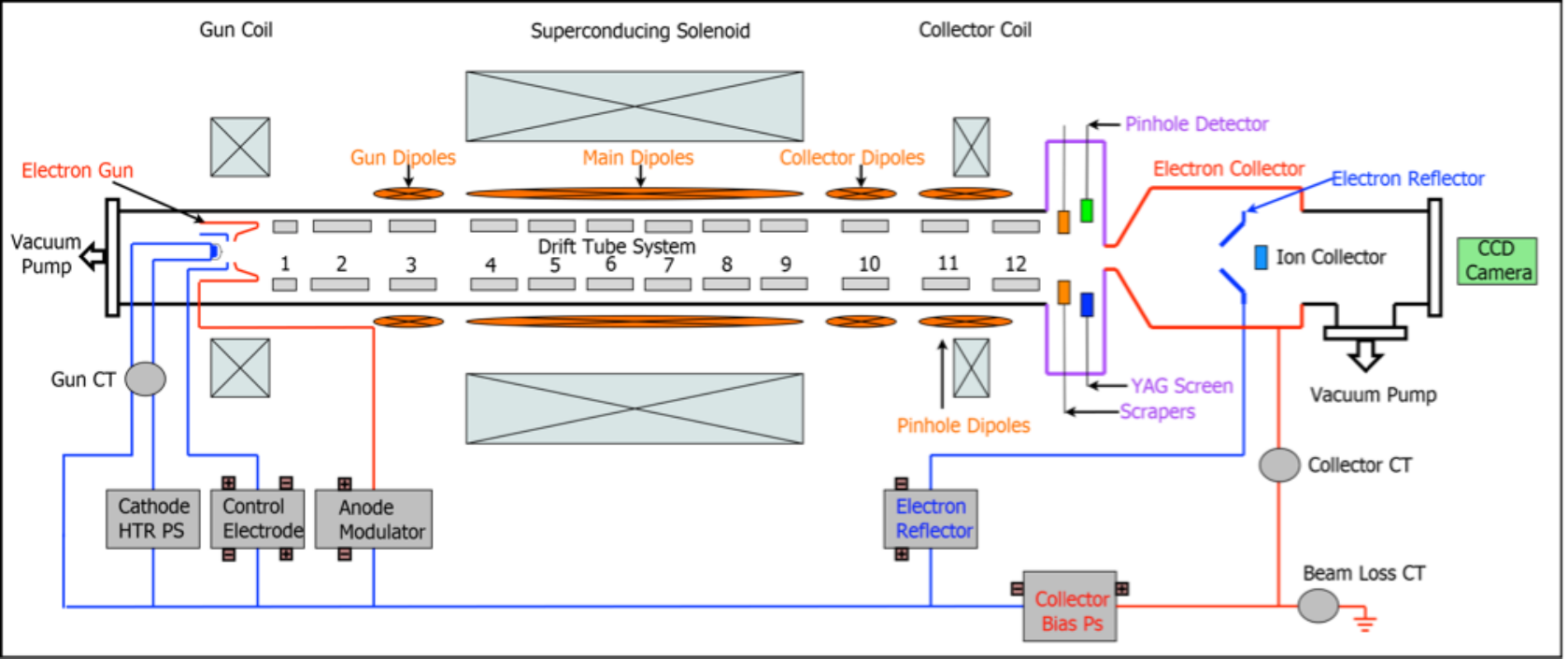}
\caption{Schematic of the electron lens test bench layout.}
\label{fig:10}
\end{figure}

\begin{figure}[htb]
\centering
\includegraphics[width=80mm]{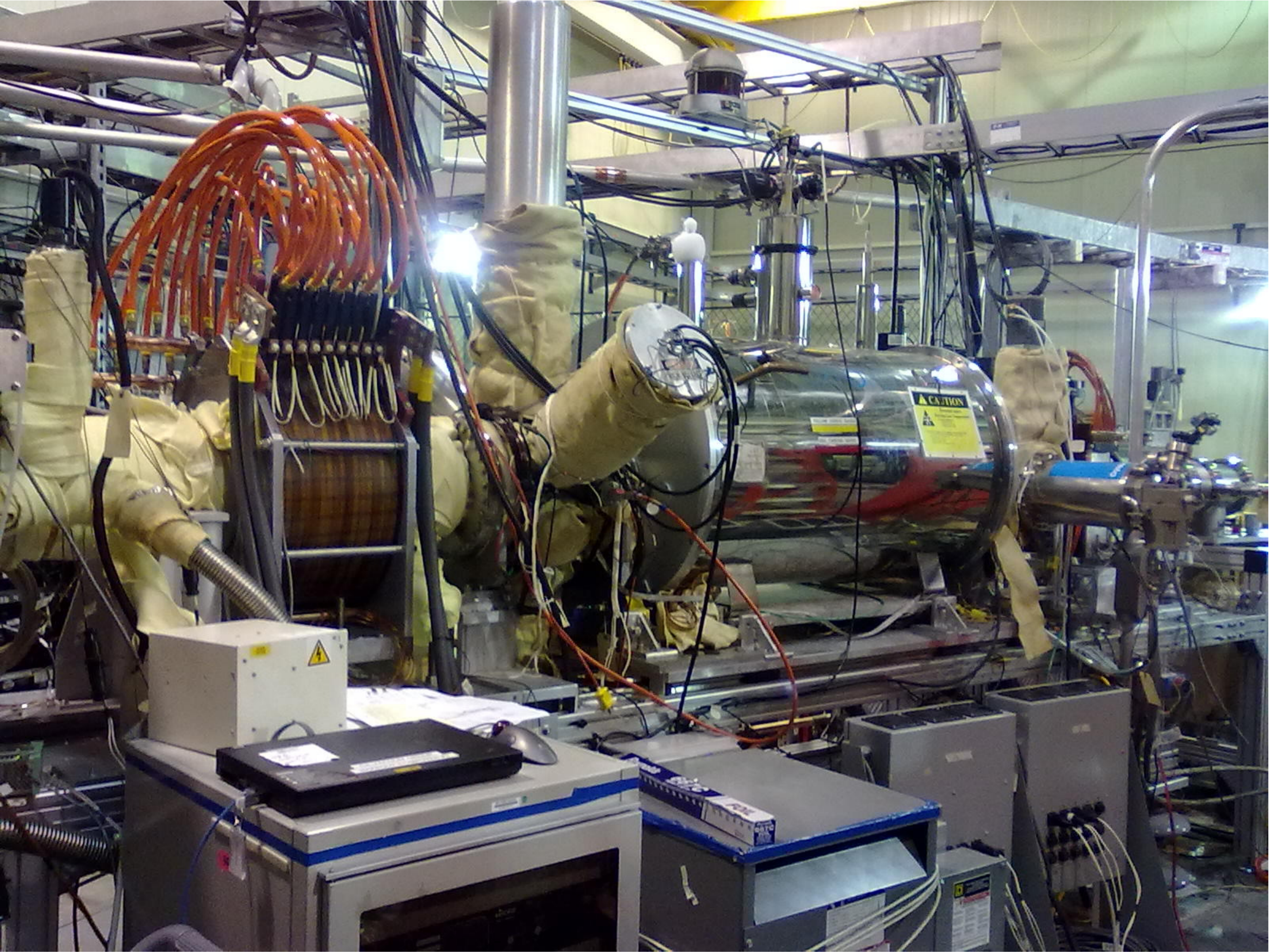}
\caption{RHIC electron lens test bench. The electron beam travels from left to
right, and GS1 is visible.}
\label{fig:2}
\vspace*{-5mm}
\end{figure}

\vspace*{-1.0mm}
The test bench (Figs.~\ref{fig:10} and \ref{fig:2}) uses the location and the
superconducting solenoid of the BNL EBIS test stand. Of the RHIC electron
lenses, the following components were installed: a gun and collector, a GS1
solenoid with power supply, a movable pin-hole detector, a movable YAG screen
with camera, and an electron halo detector.

The test bench work is complete and the following have been
demonstrated~\cite{Gu6,Gu8,Gasn,Mill2}.
\begin{itemize}
\vspace*{-3mm}
\item The gun operated in 80~kHz pulsed mode and DC mode, and reached 1~A of
  DC current with a current ripple of $\Delta I/I=0.075$\%.
\vspace*{-3mm}
\item The gun perveance with a La$_6$B cathode was measured to be
  0.93~$\mu$AV$^{-3/2}$.
\vspace*{-3mm}
\item The collector temperature and pressure was measured with the
  1~A DC current and found to be within expectations.
\vspace*{-3mm}
\item The Gaussian transverse electron beam profile was verified.
\vspace*{-3mm}
\item The machine protection system was prototyped.
\vspace*{-3mm}
\item Part of the controls software was tested.
\end{itemize}

After completion of the test bench, the components were removed and installed
in the RHIC tunnel and service building.

\section{Status and outlook}
For the ongoing RHIC Run-13, the hardware of both lenses is partially installed
(Fig.~\ref{fig:3}). The Blue lens has a complete electron beam transport
system, although instead of the superconducting main solenoid designed for the
electron lens a spare solenoid of the BNL EBIS is installed. This magnet is a
2\,m-long superconducting solenoid with a maximum field strength of 5~T, but
it does not have an iron yoke and therefore the field lines are not straight enough for
beam--beam compensation; it does, however, allow for propagation of the
electrons from the gun to the collector even at field strengths as low as 1~T. The low
field is necessary to minimize the effect on the proton spin, as long
as the second superconducting solenoid is not yet powered. The Blue lens also
has a full complement of instrumentation, with the exception of the overlap
monitor based on backscattered electrons. All drift tubes are grounded.
In this configuration, all warm magnets can be commissioned  as well as
the electron beam in pulsed mode. The two dual-plane BPMs inside the
superconducting solenoid, the YAG screen profile monitor, and the pin-hole
detector can be tested. Interaction with the proton beam is in principle
possible.

The Yellow lens has one of the new superconducting solenoids installed,
but with a straight beam pipe that does not have BPMs or drift tubes (i.e.\ the vacuum
system of the electron gun and collector is not connected to the proton beam
vacuum system). This configuration allows for commissioning of the
superconducting main solenoid and all superconducting correctors, as well as
all warm magnets. The Yellow lens is shown in Fig.~\ref{fig:9}.

The second superconducting solenoid is set up in the Superconducting Magnet
Division as a test bed for the field-straightness measurement system.
As of the submission  of this paper, the following  have been achieved.
A new lattice was commissioned for both rings that has a phase advance of
a multiple of $\pi$ between IP8 and the electron lens; for this new phase
shifter, power supplies were installed in both rings and both transverse
planes.
A bunch-by-bunch loss monitor has become available, and bunch-by-bunch BTF
measurements are being tested. The derivation of the incoherent beam--beam
tune spread in the presence of coherent modes from transverse BTF measurements
is under investigation~\cite{Goer}. In the Blue lens, a field of 1~T in the
superconducting solenoid has been established. All warm solenoids were tested
at operating currents, and all GSB and CSB solenoids ran concurrently with
RHIC polarized proton operation.

In the summer of 2013 the second superconducting main solenoid will be
installed, and the field straightness of both magnets will be measured
in place and corrected. After that, the installation will be completed for
both lenses, including the overlap detector based on backscattered electrons.

In 2014 RHIC is likely to operate predominantly with heavy ions.
The beam--beam effect with heavy ions is too small for compensation, but
all electron beam operating modes (pulsed and DC) can be established,
and the electron beam can interact with the ion beam. The first compensation
test can be done in polarized proton operation.

\section{Summary}
\vspace*{-1.0mm}
Partial head-on beam--beam compensation is being implemented in RHIC. One of
two beam--beam interactions is to be compensated with two electron lenses,
one for each of the two proton beams. This allows for an increase in the bunch
intensity with a new polarized proton source~\cite{Zele}, with the goal of
doubling the average luminosity in polarized proton operation.

The components of two electron lenses have been manufactured and partially
installed. The current installation allows for commissioning of the warm
magnets, electron beam, and instrumentation in the Blue lens. In the Yellow
lens, the new superconducting solenoid and the warm magnets can be commissioned.
First tests with ion beams are anticipated for the next year, after which
the compensation can be commissioned for polarized proton operation.

\section{Acknowledgments}
\vspace*{-1.0mm}
We are grateful to V.\ Shiltsev, A.\ Valishev, and G.\ Stancari of  FNAL for many
discussions on the Tevatron electron lenses and for the opportunities to
participate in electron lens studies at the Tevatron. We also greatly
benefited from the experience of the BNL EBIS team, including J.\ Alessi,
E.\ Beebe, M.\ Okamura, and D.\ Raparia. We had many fruitful conversations about
beam--beam and compensation problems with V.\ Shiltsev, A.\ Valishev, T.\ Sen,
and G.\ Stancari of FNAL; X.\ Buffat, R.\ DeMaria, U.\ Dorda, W.\ Herr,
J.-P.\ Koutchouk, T.\ Pieloni, F.\ Schmidt, and
F.\ Zimmerman of CERN; K.\ Ohmi of  KEK; V.\ Kamerdziev of FZ J\"{u}lich; A.\ Kabel of
SLAC; and P.\ G\"{o}rgen of TU Darmstadt. We are thankful to the U.S.\ LHC
Accelerator Research Program (LARP) for support of beam--beam simulations. We
also acknowledge the technical and administrative support received
from the BNL Superconducting Magnet Division, as well as from all groups of the
Collider-Accelerator Department, in particular K.\ Mirabella and G.\ Ganetis.

\end{document}